\documentclass[sigconf, manuscript]{acmart}
\usepackage[utf8]{inputenc}
\usepackage{graphicx}
\usepackage{amsfonts}
\usepackage{multirow, multirow}
\usepackage{hyperref}

\copyrightyear{2022}
\acmYear{2022}
\setcopyright{rightsretained}
\acmConference[FAccT '22]{2022 ACM Conference on Fairness, Accountability, and Transparency}{June 21--24, 2022}{Seoul, Republic of Korea}
\acmBooktitle{2022 ACM Conference on Fairness, Accountability, and Transparency (FAccT '22), June 21--24, 2022, Seoul, Republic of Korea}\acmDOI{10.1145/3531146.3533105}
\acmISBN{978-1-4503-9352-2/22/06}

\begin{document}
\title{De-biasing ``bias" measurement}
\author{Kristian Lum}
\email{kristianl@twitter.com}
\affiliation{%
    \institution{Twitter Inc.}
    \city{San Francisco}
    \country{USA}
}

\author{Yunfeng Zhang}
\email{yunfengz@twitter.com}
\affiliation{%
    \institution{Twitter Inc.}
    \city{San Francisco}
    \country{USA}
}

\author{Amanda Bower}
\email{abower@twitter.com}
\affiliation{%
    \institution{Twitter Inc.}
    \city{San Francisco}
    \country{USA}
}
\renewcommand{\shortauthors}{Lum, et al.}


\begin{abstract}
When a model's performance differs across socially or culturally relevant groups--like race, gender, or the intersections of many such groups--it is often called "biased." While much of the work in algorithmic fairness over the last several years has focused on developing various definitions of model fairness (the absence of group-wise model performance disparities) and eliminating such ``bias," much less work has gone into rigorously measuring it. In practice, it important to have high quality, human digestible measures of model performance disparities and associated uncertainty quantification about them that can serve as inputs into multi-faceted decision-making processes. In this paper, we show both mathematically and through simulation that many of the metrics used to measure group-wise model performance disparities are themselves statistically biased estimators of the underlying quantities they purport to represent. We argue that this can cause misleading conclusions about the relative group-wise model performance disparities along different dimensions, especially in cases where some sensitive variables consist of categories with few members. We propose the ``double-corrected" variance estimator, which provides unbiased estimates and uncertainty quantification of the variance of model performance across groups. It is conceptually simple and easily implementable without statistical software package or numerical optimization. We demonstrate the utility of this approach through simulation and show on a real dataset that while statistically biased estimators of group-wise model performance disparities indicate statistically significant differences, when accounting for statistical bias in the estimator, the estimated between-group disparities are no longer statistically significant.

\end{abstract}

\maketitle






\section{Introduction}
In the interest of developing models that behave equitably for different swaths of the population, identifying and mitigating group-wise disparities in model performance--typically measured in terms of accuracy, false positive rates, and so on--have become a central theme of creating ``fair" models. Models that exhibit group-wise performance disparities are often termed ``biased"\footnote{To be clear, we will refer to this type of ``bias'' as ``group-wise model performance disparities'' or ``performance disparities'' for short throughout our paper.}--an overloaded term, which in this case, simply means that that model performance is substantively different for some groups than others.  Canonical examples of models expressing between-group performance variability include the identification of racial disparities in false positive rates of recidivism prediction models \cite{propublica}, race-gender disparities in accuracy rates in gender classification models \cite{buolamwini2018gender}, and racial disparities in police allocation in predictive policing \cite{lum2016predict}. In each of these cases, model performance disparities were readily apparent simply by inspecting the model performance for each group separately and judging that the magnitude of difference between the groups was unacceptably large. When there are a small number of groups, each composed of many observations--as was the case in all of the above examples--this method of measuring and evaluating group-wise model performance disparities is possible. 

\begin{figure*}
    \centering
    \includegraphics[width = 5in]{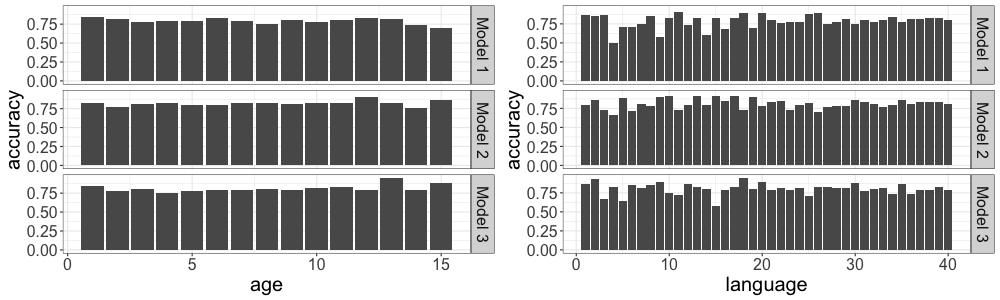}
    \caption{Each bar represents the accuracy of one of three hypothetical classification models for determining if text is spam or not broken down by age of the author of the text (left) or language of the text (right). It is relatively difficult to determine the degree of group-wise disparities for each model by only a visual examination.}
    \Description{Each bar represents the accuracy of one of three hypothetical classification models for determining if text is spam or not broken down by age of the author of the text (left) or language of the text (right). It is relatively difficult to determine the degree of group-wise disparities for each model by only a visual examination.}
    \label{fig:ex1}
\end{figure*}

When the number of groups across which performance disparities is being evaluated is large, it is difficult for a human to make judgments about the degree of disparities present simply by inspecting the model performance for all groups.  Human-understandable measurement of model performance disparities and uncertainty quantification thereof are important if decision-makers are to engage in a well-informed consideration of the trade-offs between the relative merits of a variety of candidate models, determine whether a particular model should be mitigated for performance disparities, and decide whether any model should be deployed at all. To illustrate, suppose a machine learning practitioner needs to select from among three candidate binary classification models that use text as an input. Figure \ref{fig:ex1} shows model accuracy broken down by the language of the text and age of the text's author for three hypothetical models.\footnote{Code to produce this figure, all simulations, and analysis in this paper is available at \href{https://github.com/twitter-research/double-corrected-variance-estimator}{https://github.com/twitter-research/double-corrected-variance-estimator}} The height of each bar represents each model's accuracy within the corresponding group. Which model exhibits the lowest group-wise performance disparities by age? Do any of them exhibit acceptable levels of model performance disparities with respect to age? Given a model, are there more disparities with respect to age or language? How does the group-wise performance disparities between models compare? While it is difficult to answer these questions just with three candidate models, in real-world settings, the situation can be drastically more difficult. Machine learning practitioners may be considering hundreds or thousands of models given the plethora of hyperparameter choices when training neural networks and other models. Answering these questions requires well-designed, summary metrics that transform  high-dimensional outputs of model performance on each group into lower-dimensional summaries while retaining the salient information. We call these ``meta-metrics'' since they are a metric on the vector whose coordinates contain model performance metrics on each group. We use this term to differentiate them from the ``base metric", which is the performance metric applied to each group separately .

Meta-metrics have received little attention in the algorithmic fairness literature despite their ubiquity and importance. Specifically, there have been no investigations of whether the metrics that are used to measure group-wise model performance disparities are themselves biased in the statistical estimation sense of the word. 
Uninterpretable, statistically biased meta-metrics can lead to costly outcomes. For example, suppose we obtain measurements that incorrectly suggest a model has a large degree of group-wise model performance disparities and thus requires mitigation. If we retrain the model with additional ``fairness'' constraints, we will likely end up lowering the utility of the model for every group involved unnecessarily. On the other hand, if the measurement is not easily understood, e.g. on a small scale that does not lend itself to distinguishing different scenarios,  performance disparities may go undetected. 
Unfortunately, as we show in Section \ref{sec:statBiasModelPerfDispersion}, many of the meta-metrics currently in use are statistically biased representations of the underlying quantity they purport to represent. The statistical bias arises because they fail to account for statistical uncertainty/sampling variance in the base metrics across which they are summarizing information. In particular, they are statistically biased upwards, exaggerating the extent of performance disparities, potentially leading to mitigation in cases where model performance is negligible or even identical across groups.  The statistical bias is particularly large when some sub-groups have large sampling variance, e.g. in situations where some groups consist of few individuals. This situation is common: when evaluating model performance disparities across the intersection of many group variables, as in \cite{kearns2018preventing}, even large data sets can have intersectional groups that consist of very few observations. For example, in the Adult income dataset\footnote{Downloaded from the UCI Machine Learning Repository at \url{https://archive.ics.uci.edu/ml/datasets/adult}.}-- a commonly used benchmark dataset for evaluating model performance disparities-- some intersectional groups defined by race and gender contain fewer than 50 individuals. Many intersectional groups defined by race and 10 year age bins contain only one individual per group.

Furthermore, typical approaches to quantifying statistical uncertainty of existing meta-metrics, e.g. bootstrapping, can fail to cover the true degree of between-group disparities at a much higher than expected rate because they capture uncertainty about a quantity that is itself a statistically biased estimator of the true underlying value. This failure is more pronounced when the statistical bias is large. Therefore, using existing tooling on problems with many groups, some of which are small, results in either measurements that are so statistically biased as to paint a misleading picture of between-group disparities, or using heuristics to remove small-sized groups, thereby potentially further marginalizing already marginalized groups by excluding them from group-wise model performance analysis. 



In this paper we make several contributions. First, we 
 demonstrate that commonly used meta-metrics 
depict a statistically biased representation of the true degree of model performance disparities.  As a remedy, we introduce a corrected estimator of between-group performance disparities. Second, we illustrate the dangers of a na\"ive approach to uncertainty quantification via bootstrapping for meta-metrics.  In short, we show that bootstrapping induces additional sampling variability that is unaccounted for by a simple statistical bias-corrected estimator. This leads to a situation where even the original corrected estimator (when applied to bootstrap samples) is statistically upward  biased and bootstrap intervals have very poor coverage. Third, using the case of binary classification as an example-- the most common paradigm in algorithmic fairness-- we derive a novel, conceptually simple, and easily implementable double-corrected estimator. This estimator accounts not only for the sampling variance inherent to the original base metric but also the sampling variance induced by the bootstrap procedure itself. We demonstrate that this estimator has coverage near the nominal level in simulated examples. Finally, we show using a real data example that using na\"ive methods for quantifying model performance disparities can indicate substantial disparities in model performance whereas using our estimator that appropriately accounts for statistical uncertainty in the base metrics leads to the opposite conclusion. Taken in whole, this work illustrates the delicate and difficult nature of rigorously measuring and assessing uncertainty about the between-group model performance disparities that have come to characterize the field of algorithmic fairness. 

The paper proceeds as follows. In Section \ref{sec:relatedWork}, we survey related work on measuring model performance disparities. In Section \ref{sec:notation}, we introduce notation and in Section \ref{sec:statBiasModelPerfDispersion} we show mathematically and through simulation that all meta-metrics we have identified suffer from statistical bias. In Section \ref{sec:doubleCorrected}, we develop a statistically unbiased estimator for summarizing group performance disparities with associated uncertainty quantification, along the way illustrating the pitfalls with na\"ive application of meta-metrics and bootstrapping. In Section \ref{sec:realData}, we present our example using the Adult Income dataset. In Section \ref{sec:conclusion}, we close with recommendations and pointers to corners of the statistics literature that address very similar problems that will likely prove useful to this task.

\section{Related Work}\label{sec:relatedWork}


Many group fairness papers assume that there are only two demographic groups of interest--a privileged group and a non-privileged group--in order to make theory feasible or presentation digestible \cite{liu2018delayed,hu2019disparate, bakalar2021fairness, zehlike2017fa, diciccio2020evaluating, friedler2019comparative, yan2020silva}. Although assessing binary group disparities in this case is straight forward, e.g. using the absolute value of the difference in false positive rates between two groups as a measure of unfairness \cite{bechavod2017penalizing}, this assumption is not appropriate in many real-world settings. While there are some papers that introduce theory or methods that can handle arbitrarily many demographic groups, many of these papers do not synthesize group disparities, and some even turn to binary demographic groups when validating their approaches \cite{hashimoto2018fairness, zhang2018mitigating, madras2018learning, dwork2018decoupled, gorantla2021problem}. Furthermore, empirical work typically shows bar graphs or tables to illustrate model performance disparities over multiple demographic groups. Although this granularity of information is undoubtedly useful and important on its own, these works do not provide a way to quickly synthesize group disparities of model performance across many potential models \cite{buolamwini2018gender, zehlike2022fair, huang2020multilingual,adragna2020fairness}. 
 
Outside of the typical group fairness literature, there have been several other threads that address model performance disparities. While individual fairness approaches \cite{dwork2012fairness} provide a way to measure disparities in a group-free way--seemingly alleviating the problem we set out to solve--oftentimes to practically validate these individual fairness approaches, group-wise disparities are measured \cite{yurochkin2019training, bower2021individually} leading us back to our initial problem. Second, statistical dependence between demographic groups and model predictions is a common way to measure group-wise model performance disparities \cite{komiyama2018nonconvex, mary2019fairness, tramer2017fairtest}, e.g. correlation. However, the degree of dependence does not tell us the degree of group disparities in model performance, so these approaches in general are inappropriate for our purposes. In addition, we consider categorical demographic groups, not continuous demographic groups, so many of these approaches are not even applicable for the setting we study. Third, economic inequality metrics have been proposed for summarizing model disparities at the individual level without demographic information \cite{lazovich2021measuring, speicher2018unified, saint2020fairness, mccurley2008income, bandy2021more}. While undoubtedly important, these demographic-free measures of inequality in the allocation of benefits from a model address a different problem. Demographic disparities are in and of themselves an important component of understanding the differential impacts of a model.



 There are a small number of papers that propose approaches for practically measuring disparities when there are many demographic groups \cite{agarwal2018reductions, kearns2018preventing, ghosh2021characterizing, geyik2019fairness, johndrow2019algorithm}. All of these papers synthesize group disparities by either reporting the the maximal deviation from average performance (sometimes normalized by subgroup size \cite{kearns2018preventing}), the mean absolute deviation from average performance \cite{johndrow2019algorithm}, or some measure of the disparity between the maximum and minimum of the base metrics. 

Furthermore, there is a growing collection of open source software tools for measuring and mitigating group-wise model performance disparities of machine learning models such as IBM’s AI Fairness 360 \cite{bellamy2018ai}, Microsoft’s FairLearn \cite{bird2020fairlearn}, Google’s TensorFlow Fairness Indicators \cite{google}, LinkedIn’s LiFT \cite{vasudevan2020lift}, FairTest \cite{tramer2017fairtest}, Aequitas \cite{saleiro2018aequitas}, FAT Forensics \cite{sokol2019fat}, FairVis \cite{cabrera2019fairvis}, FairSight \cite{ahn2019fairsight}, Themis \cite{angell2018themis}, Silva \cite{yan2020silva}, and Responsibly \cite{responsibly}. Most of these tools can only handle binary demographic groups.  For those that can handle arbitrarily many groups, they tend to use simple meta-metrics without consideration of statistical bias or cannot aggregate group disparities. IBM’s AI Fairness 360 tool \cite{bellamy2018ai} synthesizes between-group disparities via the generalized entropy index, although the majority of their group fairness metrics can only handle two demographic groups. Although Google TensorFlow Fairness Indicators tool can handle arbitrarily many demographic groups, users are required to determine the disparities through bar charts or tabular data that contains the measurements on each of the groups. Their approach is very similar to the hypothetical example depicted in Figure \ref{fig:ex1}. Microsoft’s FairLearn tool \cite{bird2020fairlearn} can handle arbitrarily many groups where group disparities are measured as a function of the maximum and minimum of model performance over all the groups. 
LinkedIn's LiFT tool \cite{vasudevan2020lift} measures variability in group disparities through economic inequality metrics like the generalized entropy index. 
Neither FairLearn nor LiFT quantifies the uncertainty of the point estimates of these group disparity measures.

Table \ref{metametric-table} summarizes the meta-metrics we have identified in both the academic literature and in open-source tools for group-wise model performance disparities measurement and mitigation. We  include one additional meta-metric that we have not found used to summarize disparities in the literature or open source tools: the variance. Our reason for including the variance will quickly become apparent.


\begin{table*}
\begin{center}
\begin{tabular}{ l |l | l | l }
 Meta-Metric Name & Formula & Type & Used by\\ \hline
 max-min difference & $M_{\text{mm-diff}}(Y) = \max_k Y_k - \min_k Y_k$ & Extremum & \cite{bird2020fairlearn} \\  
 max-min ratio & $M_{\text{mm-ratio}}(Y) = \frac{\max_k Y_k}{\min_k Y_k}$  & Extremum & \cite{bird2020fairlearn, ghosh2021characterizing} \\
 max absolute difference & $M_{\text{max-abs-diff}}(Y) = \max_k |Y_k - \bar{Y}|$ & Extremum & \cite{kearns2018preventing, agarwal2018reductions}\\
 mean absolute deviation & $M_{\text{mad}}(Y) = \frac{1}{K}\sum_k |Y_k - \bar{Y}|$ & Variability &  \cite{johndrow2019algorithm}\\
 variance & $M_{\text{var}}(Y) = \frac{1}{K-1} \sum_k (Y_k - \bar{Y})^2$ & Variability & \\
 generalized entropy index ($\alpha \neq 0,1$) & $\frac{1}{K \alpha (\alpha-1)}\sum_{k=1}^K\left[\left(\frac{Y_k}{\overline{Y}}\right)^\alpha - 1\right]$ & Variability & \cite{bellamy2018ai, vasudevan2020lift, speicher2018unified} \\ 
\end{tabular}
\end{center}
\caption{\label{metametric-table} Meta-metrics for summarizing between-group variability in model performance. }
\Description{Meta-metrics for summarizing between-group variability in model performance.}
\end{table*}

\section{Notation}\label{sec:notation}
Suppose we have $n$ individuals, where associated with each individual is a covariate vector $x$ and an outcome variable $w$. We use model $f(x)$ to predict $w$. These individuals fall into $K$ different groups, which we index by $k = 1, ..., K$, each consisting of $n_k$ individuals. Because we will not be dealing further with observations at the individual level, we define $w_k$ and $x_k$ to be the set of outcomes and covariate vectors, respectively, for all individuals who fall in group $k$, and overload $f(x_k)$ to mean $f$ applied to each covariate vector in $x_k$ individually. In the fairness literature, the groups usually denote different demographic groups or other salient social characteristics across which one might desire comparable model performance. Common choices include race, gender, location, and intersections thereof.  To each group, we apply a base model performance metric, $m$, that summarizes how well the model performs for individuals within that group. We define $Y_k := m(w_k, f(x_k))$ to be the observed model performance for group $k$. Unlike in conventional modeling settings where $Y$ typically represents an observation for one individual, $Y_k$ represents a summary statistic of model performance across all $n_k$ individuals who fall in group $k$. $Y_k$ is itself a statistic with ``true" model performance $\mu_k = \mathbb{E}[Y_k]$ and standard error $\sigma_k$. To be clear, $\sigma_k^2$ is the sampling variance of the {\it estimator} $Y_k$, not the variance of the data itself. Finally, let $Y:= [Y_1, \dots, Y_k]^T$ and $\mu = [\mu_1, \dots, \mu_k]^T$ be $k$-dimensional vectors, with $\bar{Y}$ and $\bar{\mu}$ the mean of the vectors, respectively.

As in much of the algorithmic fairness literature, our examples focus on binary classification. 
In this setting, common choices of the base metric, $m$, are the accuracy, false positive rate, positive predictive value, etc. In all examples presented, we simulate from the following model

\begin{eqnarray}\label{eq:sim-model}
Z_k \sim \text{Binomial}(n_k, \mu_k) &  &
Y_k  =  \frac{Z_k}{n_k}
\end{eqnarray}

This simulation framework is flexible enough to accommodate many of the base metrics that typically would be used for binary classification, though the interpretation of $n_k$ and $Z_k$ is metric-dependent. For example, if $m$ is accuracy, then $Z_k$ represents the number of correct classifications and $n_k$ represents the total number of individuals in the group. If $m$ is the false positive rate, then $Z_k$ is the number of negative classifications among those observations that were truly positive and $n_k$ is the number of positives in group $k$.

\section{Statistical bias in group-wise model performance disparities metrics}\label{sec:statBiasModelPerfDispersion}
Whereas the metric $Y_k = m(w_k, f(x_k))$ is applied to the data and predictions {\it within} a group to calculate model performance for that group specifically, meta-metrics summarize the degree to which the model's performance varies across groups. These summary measures, $M(Y)$, take in the vector of observed model performance, $Y = [Y_1, ..., Y_K]$, and output a number that measures the degree of group-wise model performance disparities, i.e. a distance from the ideal state in which the model performs equally well for all groups. Here we consider statistical bias (the expected difference between the estimator and the truth) in several metrics that have been used in the algorithmic fairness literature, have been implemented in open source software for measuring and remediating disparities, and/or are common statistical measures of variability. Table \ref{metametric-table} defines several such metrics.

Suppose we are interested in measuring the ``true" degree to which model performance varies across groups, $M(\mu)$. We don't observe $\mu$, but we do observe $Y$-- a vector of statistically unbiased estimates of $\mu$. What happens if we use $M(Y)$ as a measurement of $M(\mu)$? For several of the considered metrics, it is easy to see mathematically that $M(Y)$ is a  statistically biased estimate of $M(\mu)$, though an expression for the expected upward-deviation is not available. For example, for the mean absolute deviation, 

\begin{eqnarray*}
    \mathbb{E}[M_{\text{mad}}(Y)] &= \mathbb{E}\left[\frac{1}{K} \sum_{k=1}^K |Y_k - \bar{Y}|\right] \\
    &= \frac{1}{K}  \sum_{k=1}^K \mathbb{E} \left| Y_k - \bar{Y} \right|  \\
    & > \frac{1}{K} \sum_{k=1}^K  \left|\mathbb{E} \left[Y_k - \bar{Y}  \right] \right| \\
    & = \frac{1}{K} \sum_{k=1}^K |\mu_k - \bar{\mu}|  \\
    & = M_{\text{mad}}(\mu),
\end{eqnarray*}
where the inequality comes by applying Jensen's inequality to each term in the sum since the absolute value is a convex function. Similar arguments relying on Jensen's inequality, can be made to show that $M_{\text{max-abs-diff}}(Y)$ is also a positively statistically biased estimator of $M_{\text{max-abs-diff}}(\mu)$. 

For the variance, we can get an analytical expression for the expectation of $M_{\text{var}}(Y)$ in terms of $M_{\text{var}}(\mu)$ and the $\sigma_k$s: 

\begin{eqnarray*}
\mathbb{E}[M_{\text{var}}(Y)] & = &  \mathbb{E} \left [\frac{1}{K-1}\sum_k ( Y_k - \bar{ Y})^2 \right ] \\
& = & M_{\text{var}}(\mu) +  \left( \frac{1}{K}\sum_k \sigma_k^2 \right),
\end{eqnarray*}

\noindent This is derivable using the fact that $\mathbb{E}[{X^2}] = \mathbb{E}[{X}]^2 + \mathrm{Var}(X)$ and $\sum_k (x_k - \bar{x})^2 = \sum_{k=1}^K x_k^2 - K \bar{x}^2$. The derivation is given in the appendix. Because $\sigma_k^2 \geq 0$, this shows that $M_{\text{var}}(Y)$ is also an upwardly statistically biased estimator of $M_{\text{var}}(\mu)$, the ``true" between-group model performance variance. 

\begin{figure*}
    \centering
    \includegraphics[width = 6in]{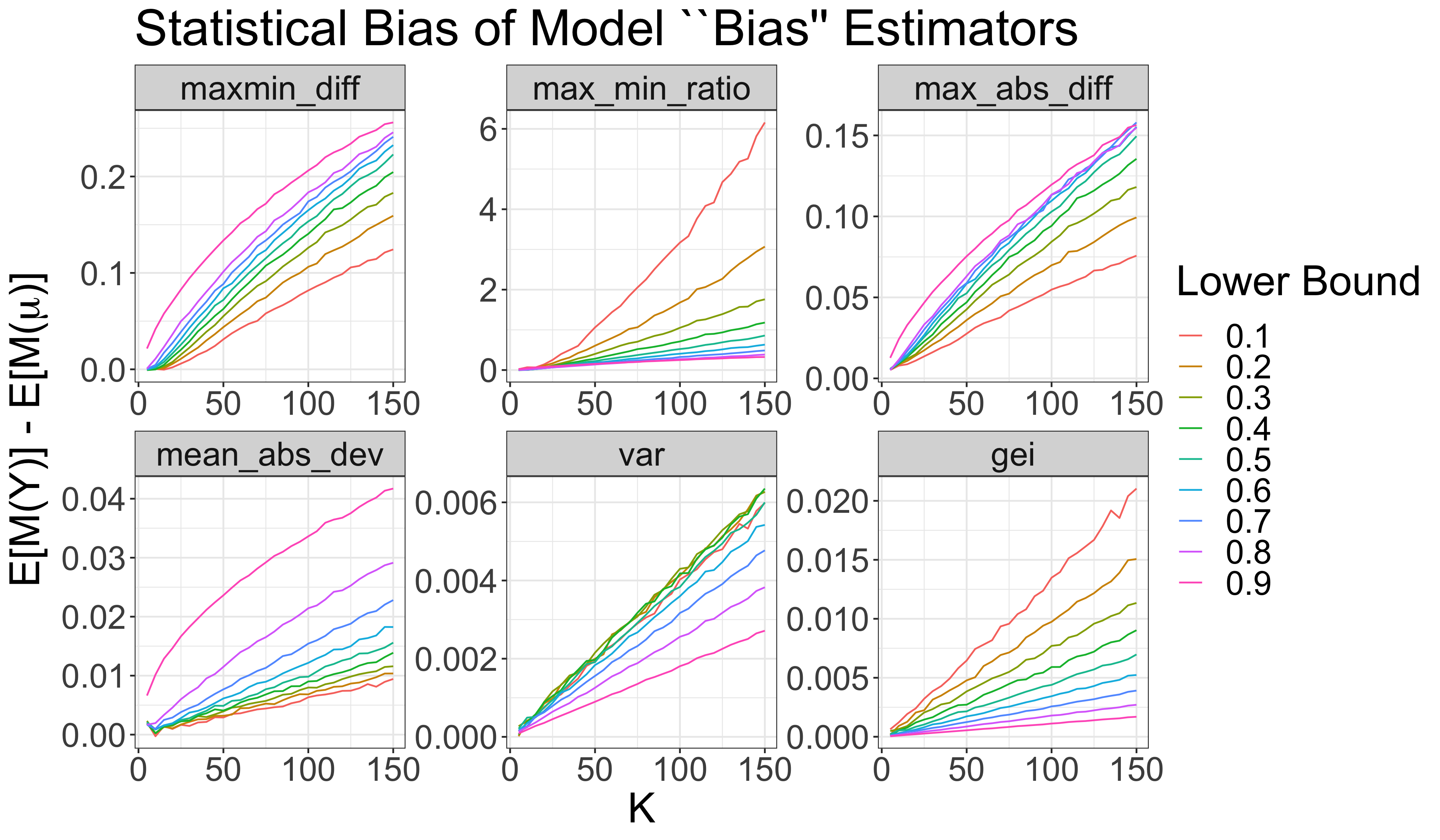}
    \caption{Empirical illustration that na\"ive estimates of meta-metrics in Table \ref{metametric-table} are statistically biased upwards. $K$ on the $x$-axis indicates the number of groups. The sum of the number of samples over all the groups is constant regardless of $K$ and the number of samples per group is equal for any given $K$.}
    \Description{Empirical illustration that na\"ive estimates of meta-metrics in Table \ref{metametric-table} are statistically biased upwards. $K$ on the $x$-axis indicates the number of groups. The sum of the number of samples over all the groups is constant regardless of $K$ and the number of samples per group is equal for any given $K$.}
    \label{fig:maxmin}
\end{figure*}

We have shown mathematically that several of the meta-metrics are statistically biased. Others are not so mathematically tractable. We investigate all meta-metrics in Table \ref{metametric-table} empirically through simulation. In our simulation, five thousand individuals are evenly divided into $K$ groups for $K$ ranging from five to 150. That is, $n_k = \text{round}\left (\frac{5,000}{K} \right )$. In cases where $K$ does not evenly divide $n$, the total population size deviates slightly from 5,000 due to rounding. We set $\mu$ to be the corresponding vector of $K$ ``true" model performance values equally spaced on $[l, .9]$ (i.e. $\mu_k = l + \frac{k-1}{K-1}(.9 - l)$) and consider several different values of $l$, the lower bound. Given $\mu_k$ and $n_k$, we simulate from the model described in \eqref{eq:sim-model}. Figure \ref{fig:maxmin} shows a Monte Carlo estimate of the  statistical bias, $M(Y) - M(\mu)$, averaged over 1,000 simulations for several values of $K$ and $l$, for each of the considered meta-metrics. A different meta-metric is given in each panel of the figure.

For all considered meta-metrics, $M(Y)$ over-estimates $M(\mu)$. In practice, this is problematic for two reasons. In an absolute sense, this can paint a misleading picture of the extent to which a model performs differently across groups. For decision-makers who need this information to determine how and whether a model should be used or modified, exaggerating the degree to which the model is ``unfair" may cause unnecessary model adjustments. In the context of the well-known fairness-accuracy trade-off \cite{dutta2020there}, having misleading measures of group-wise model performance disparities may cause model builders to trade off more accuracy than necessary in the name of eliminating group-wise model performance disparities, resulting in a less performant model for all individuals. In a relative sense, this can lead to misleading conclusions about how group-wise model performance disparities compares along different axes. For example, suppose we wanted to evaluate the degree of disparities for two ``sensitive variables": age and language, as in the example in Figure \ref{fig:ex1}. These variables have different numbers of groups, with $K=15$ for age and $K=40$ for language. In the case where individuals are evenly distributed across groups as in this simulation, we can see that all of our metrics would, on expectation, return a higher estimate of group-wise model performance disparities for the variable with $K=100$ than the variable with $K=10$, all else held constant. That is, we would erroneously conclude that the model had more group-wise model performance disparities  with respect to language than age, even when the true between-group variability in model performance was identical for both variables. This remains true even when the model is ``fair" for both axes (i.e. when the lower bound is 0.9 and all groups have equal true performance). 

In this simulated example, the statistical bias increases as a function of $K$ for all metrics. This is an artifact of the fact that the observations are equally distributed across the groups. Comparing the statistical bias of a ``sensitive variable" with individuals evenly distributed across many groups and another ``sensitive variable" with individuals unevenly distributed across fewer groups may not exhibit the same pattern of a larger statistical bias for the many-group variable than the few-group variable. 

\section{Correcting for statistical bias in group-wise model performance disparities measurement}\label{sec:doubleCorrected}

Given the statistical bias in all of the existing meta-metrics, what is to be done? The expression for the statistical bias in the variance offers one possibility. Because we know the statistical bias in $M_{\text{var}}(Y)$ as a function of the $\sigma_k$s, we can correct for it and use the estimator

\begin{equation}
\label{eq:ho}
\hat M_{\text{var}}(\mu) = M_{\text{var}}(Y) - \frac{1}{K} \sum_k \hat \sigma_k^2, 
\end{equation}

\noindent where $\hat \sigma_k$ is the standard error of the estimate of model performance for group $k$.  As it turns out, \eqref{eq:ho} was first introduced by Cochran in the random effects ANOVA context in 1954 \cite{cochran1954combination} and was popularized for meta-analysis of between-study variability by Hedges and Olkin in 1985 \cite{hedges2014statistical}. Often, in these other contexts, this estimator is truncated to preclude the possibility of a negative variance estimate, i.e.
$\hat M_{\text{var}}(\mu) = \max(0, M_{\text{var}}(Y) - \frac{1}{K} \sum_k \hat \sigma_k^2)$. We also adopt this convention. The estimator given in Equation \eqref{eq:ho} offers a conceptually clean and easily calculable estimate of between-group model performance variance. The untruncated version is statistically unbiased when statistically unbiased estimates of the standard errors are also available \cite{viechtbauer2005bias}.  We find this estimator appealing because it is easy to explain and easy to calculate without access to statistics-specific software packages. Going forward, due to the mathematical tractability of the variance as a meta-metric, we focus our investigation on measuring and quantifying uncertainty about the variance of model performance across groups. 

\subsection{Simulation example}
We extend the simulation framework to focus in on four scenarios. These scenarios are summarized in Table \ref{tab:scenarios}. In all scenarios presented, we consider a training set of size of $n = 5000$ with $K = 100$. In the Equal Group-size condition, all groups have $\frac{n}{K}  = 50 $ observations. In the Unequal Group-size condition, we create group sizes by linearly interpolating between 10 and 90 and rounding. The minimum group size is 10 and the maximum 90. Every integer between 10 and 90 has at least one group of that size; occasionally due to rounding there are two groups with the same number of observations. In the Equal Performance condition, $\mu_1 = \mu_2 = ... = \mu_k = 0.8$. In these scenarios, the true variance across groups is zero. In the Unequal Performance condition, group performance is equally spaced between 0.1 and 0.9, as in the simulation in Figure \ref{fig:maxmin}. Here, the true variance in performance across groups is 0.055. For each scenario, we again use the simulation model given in \eqref{eq:sim-model}. 

\begin{table*}
\begin{center}
\begin{tabular}{ c |c | c }
 & Equal Group-size & Unequal Group-size \\ \hline \hline
  \multirow{2}{*}{Equal Performance} &  $\mu_1 = \mu_2 = ... = \mu_k = 0.8$ & $\mu_1 = \mu_2 = ... = \mu_k = 0.8$\\
  & $n_k = \frac{n}{K}$ & $n_k = \text{round}(10 + 80 \frac{k-1}{K-1})$ \\ \hline
  \multirow{2}{*}{Unequal Performance} &  $\mu_k = .1 + 0.8 \frac{k-1}{K-1}$ & $\mu_k = .1 + 0.8 \frac{k-1}{K-1}$\\
  & $n_k = \frac{n}{K}$ & $n_k = \text{round}( 10 + 80 \frac{k-1}{K-1})$ \\ \hline
\end{tabular}
\end{center}
\caption{\label{tab:scenarios} The four scenarios used for simulations. We consider the case when there are no group-wise model performance disparities (equal performance) or not (unequal performance) and also whether the number of samples in each group is equal or not.}
\Description{\label{tab:scenarios} The four scenarios used for simulations. We consider the case when there are no group-wise model performance disparities (equal performance) or not (unequal performance) and also whether the number of samples in each group is equal or not.}
\end{table*}

For each scenario, we simulate data and calculate the variance of $Y$ and the corrected variance of $Y$ as in \eqref{eq:ho}. To estimate $\hat \sigma^2$, we use the plug-in estimator of the sampling variance of a binomial proportion, $\hat \sigma^2 = \frac{Y ( 1-Y)}{n_k}$. We repeat this 1000 times. Figure \ref{fig:experiment1} shows the distribution of these estimates. While we have already seen (both mathematically and through simulation) that the uncorrected variance (red) is upwardly statistically biased, this shows that correcting for the statistical bias by substracting off an estimate of the average standard error of the groups successfully results in between-group variance estimates that are centered at the truth (vertical black line), i.e. statistically unbiased. Next we explore what happens if we bootstrap this corrected variance estimator to obtain bootstrap intervals for uncertainty quantification. 

\begin{figure*}[h]
\includegraphics[width = 5in]{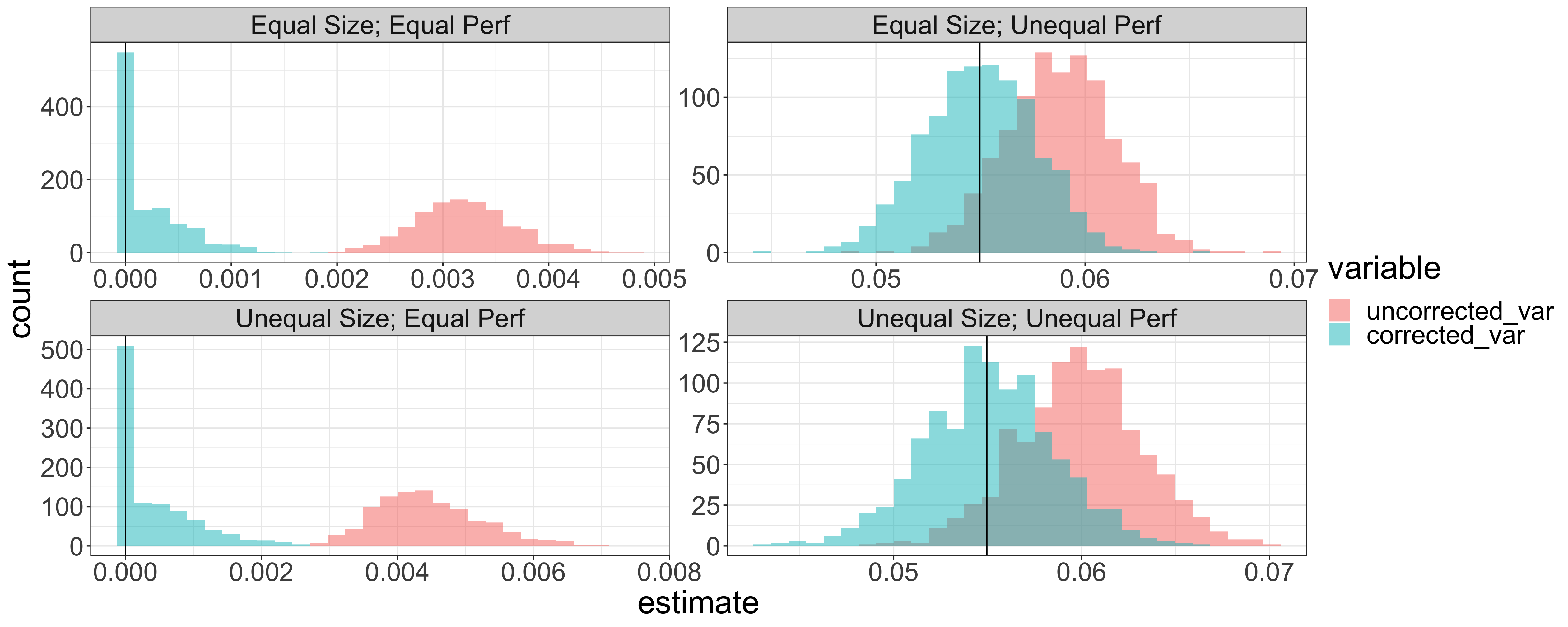}
\caption{Each panel shows a different simulation scenario depending on whether the group sizes are equal or unequal and whether there are truly disparities or not in model performance among the groups. The black vertical lines indicate the true variance of model performance among the groups. We show the histogram of uncorrected (red) and corrected (blue) estimates of between-group variability in model performance across 1000 simulated replicates. The uncorrected estimate is statistically biased upwards.  \label{fig:experiment1}}
\Description{Each panel shows a different simulation scenario depending on whether the group sizes are equal or unequal and whether there are truly disparities or not in model performance among the groups. The black vertical lines indicate the true variance of model performance among the groups. We show the histogram of uncorrected (red) and corrected (blue) estimates of between-group variability in model performance across 1000 simulated replicates. The uncorrected estimate is statistically biased upwards.  \label{fig:experiment1}}
\end{figure*}


\subsection{Bootstrapping the (statistical) bias-corrected variance estimate}\label{sec:bootstrap}
The corrected variance estimator defines an easy and explainable way to obtain a statistically unbiased point estimate of the true between-group performance variance. Bootstrapping offers a similarly conceptually simple and easily implementable method for calculating uncertainty intervals. To create non-parametric bootstrap intervals, one samples  with replacement from the observed data. If we consider the number of individuals in each group to be fixed (as we do in the following examples), then for each group $k$, we sample $n_k$ individuals with replacement from the collection of all individuals in group $k$. This makes one bootstrap sample, $Y^* = [Y_1^*, ..., Y_K^*]$. For each bootstrap sample, a statistic-- such as the corrected variance-- is calculated. This process is repeated $B$  times, resulting in $B$ samples of the target statistic. Uncertainty intervals for the statistic are calculated as the empirical quantiles of the $B$ estimates. For example, a standard 95\% interval would be calculated by setting the lower and upper bounds of the interval to be the 0.025 and 0.975 quantiles of the samples of the statistic, respectively. There are many techniques for bootstrapping, such as versions that use the bootstrap samples only to calculate standard errors around point estimate obtained from the original sample \cite{efron1994introduction}. Here, we use the ``percentile" bootstrap. Future work should consider other techniques for bootstrapping in this context, such as using single-corrected estimator as a point estimate and the double-corrected estimator to obtain intervals about that point estimate.

\begin{figure*}
    \centering
    \includegraphics[width = 5in]{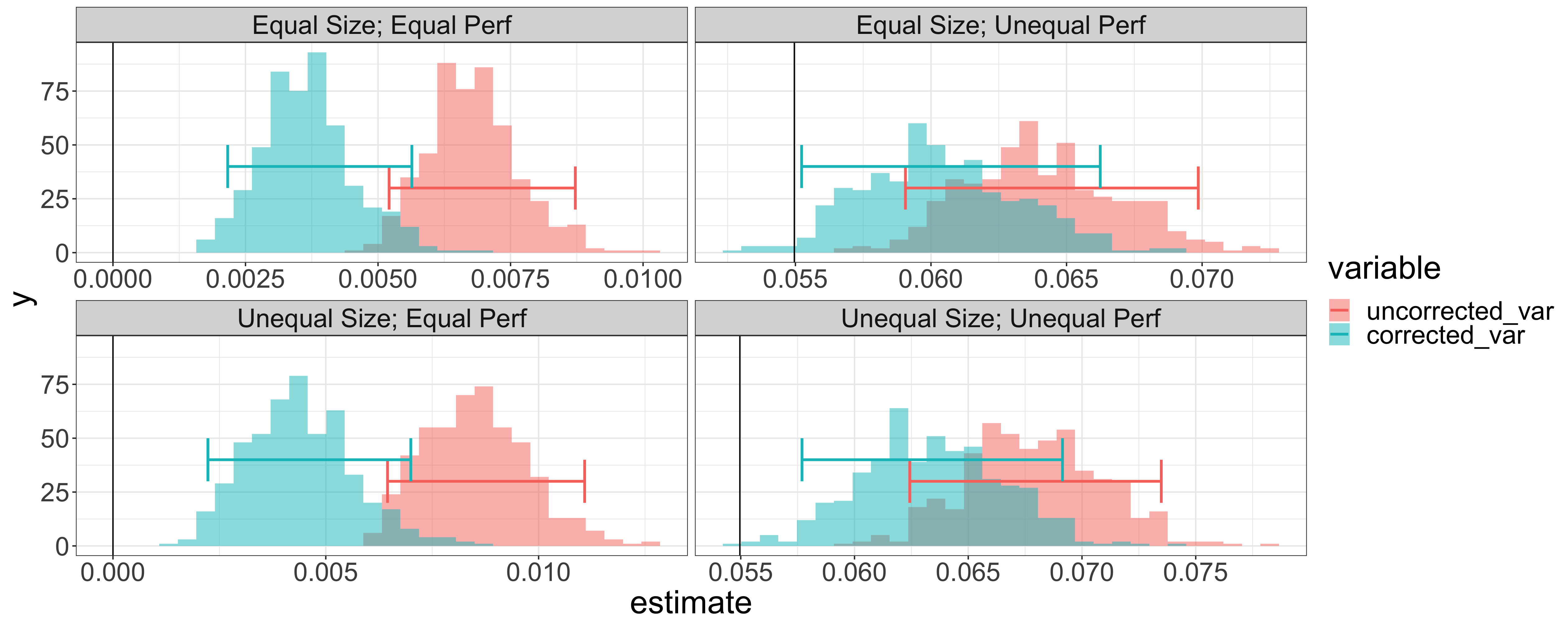}
    \caption{ Each panel shows a different simulation scenario depending on whether the group sizes are equal or unequal and whether there are truly disparities or not in model performance among the groups. We show the histogram of bootstrap samples of the corrected (blue) and uncorrected (red) variance for \textit{one draw from the generative model}. Intervals represent 95\% bootstrap intervals. Neither estimator covers the true variance of model performance among the groups shown by the vertical black lines.} \label{fig:experiment2}
    \Description{ Each panel shows a different simulation scenario depending on whether the group sizes are equal or unequal and whether there are truly disparities or not in model performance among the groups. We show the histogram of bootstrap samples of the corrected (blue) and uncorrected (red) variance for \textit{one draw from the generative model}. Intervals represent 95\% bootstrap intervals. Neither estimator covers the true variance of model performance among the groups shown by the vertical black lines.}
\end{figure*}

Figure \ref{fig:experiment2} shows the distribution of bootstrap samples of the corrected and uncorrected variance for one draw from the generative model. The horizontal bars show 95\% bootstrap intervals for the corrected variance. It is important to differentiate that whereas in Figure \ref{fig:experiment1}, the histogram is made up of single estimates of the variance each corresponding to independent draws from the generative model, in Figure \ref{fig:experiment2} the histogram is generated by bootstrap re-sampling from {\it one} draw from the generative model to obtain uncertainty estimates for the  variance estimate associated with that single draw. Here, we see that the na\"ive application of the (corrected) variance estimator to each bootstrap sample results in distributions of estimates that are systematically shifted upwards with intervals (horizontal lines) that do not even come close to covering the true value (shown by the vertical black line). When bootstrapping, even our statistically unbiased estimator of the variance is statistically biased. 

\subsection{What happened?}
Let's go back to our original model given in \eqref{eq:sim-model}. Under this model, the variance of our estimator $Y_k$ is given by $\hat \sigma^2_k = \frac{Y_k (1-Y_k)}{n_k}$, which is what we've used as our estimate of the sampling variance of $Y_k$ when applying the correction. However, when we generate bootstrap samples (by re-sampling the observations within groups), our generative model for each bootstrap sample becomes the following:

\begin{eqnarray*}
X_k  \sim \text{Binomial}(n_k, \mu_k)
& & Y_k  = \frac{X_k}{n_k} \\
X^*_k  \sim \text{Binomial}(n_k, Y_k) & & 
Y^*_k  = \frac{X^*_k}{n_k}
\end{eqnarray*}

Applying iterated expectations, we find that the sampling variance of each bootstrap sample is given by 

\begin{eqnarray*}
\mathrm{Var}(Y_k^*) & = & \mathbb{E}(\mathrm{Var}(Y^*_k \mid Y_k)) + \mathrm{Var}(\mathbb{E}(Y_K^* \mid Y_K)) \\
& = & \mathbb{E}\left(\frac{Y_k ( 1 - Y_k)}{n_k}\right) + \mathrm{Var}(Y_k) \\
& = & \mathbb{E}\left (\frac{Y_k}{n_k} - \frac{Y_k^2}{n_k}\right) + \frac{\mu_k(1-\mu_k)}{n_k}\\
& = & \frac{\mu_k}{n_k} - \left (\frac{\mathrm{Var}(Y_k)}{n_k} + \frac{\mathbb{E}(Y_k)^2}{n_k} \right) + \frac{\mu_k(1-\mu_k)}{n_k}\\
& = & \frac{\mu_k}{n_k} - \frac{\mu_k(1-\mu_k)}{n_k^2} - \frac{\mu_k^2}{n_k} + \frac{\mu_k(1-\mu_k)}{n_k} \\
& = & \frac{2 \mu_k(1-\mu_k)}{n_k} - \frac{\mu_k(1-\mu_k)}{n_k^2}
\end{eqnarray*}

Again, using a plug-in estimate of the sampling variance of $Y_k^*$, we get $\hat \sigma^{2*}_k = \frac{2  Y^*_k(1-Y^*_k)}{n_k} - \frac{Y^*_k(1-Y^*_k)}{n_k^2}$. This implies that for each bootstrap sample, a ``double-corrected" estimate of the variance of $\mu$ is given by

$$ \hat {M}_{\text{var}}(\mu) = M_{\text{var}}(Y^*) - \frac{1}{K}\sum_k \hat \sigma^{2*}_k.$$

\begin{figure*}
    \centering 
    \includegraphics[width = 5in]{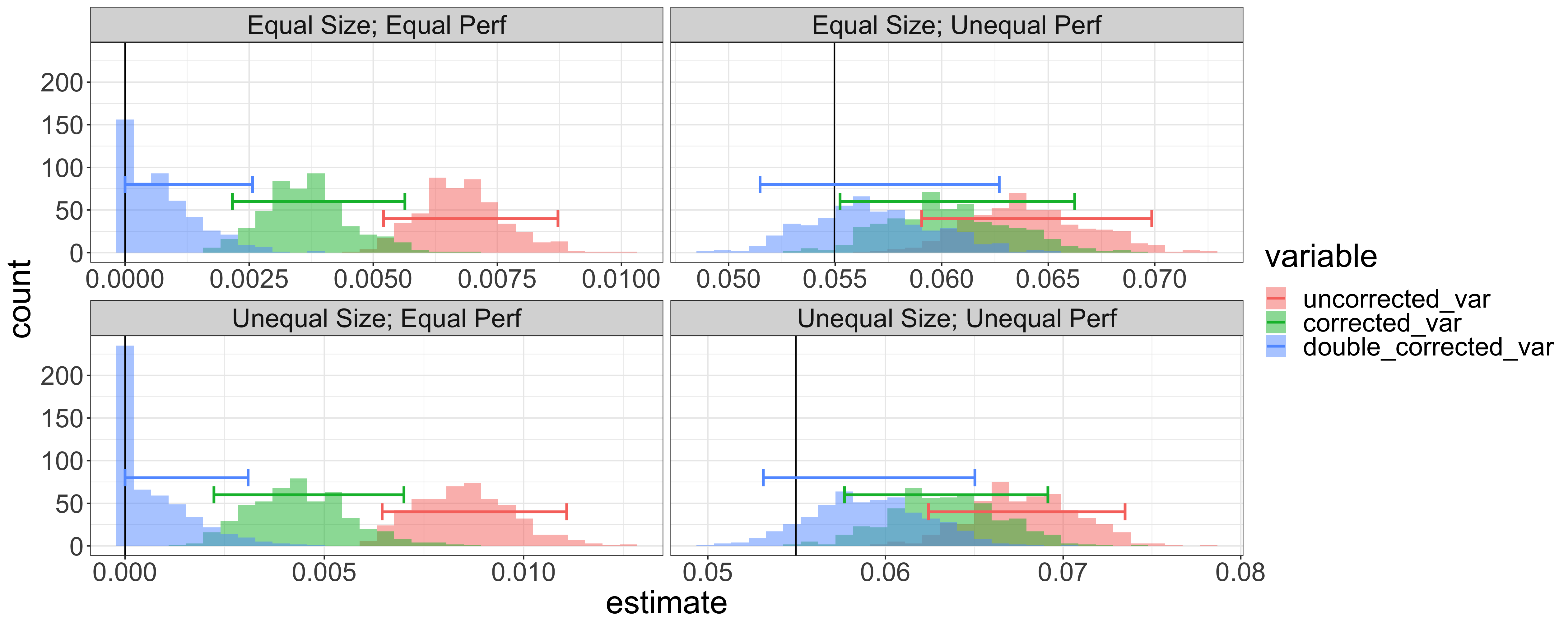}
    \caption{Each panel shows a different simulation scenario depending on whether the group sizes are equal or unequal and whether there are truly disparities or not in model performance among the groups. We show the histogram of 500 bootstrap samples and 95\% bootstrap intervals for the uncorrected (red), corrected (green), and double-corrected (blue) estimators of the between-group variance for one draw from the generative model.  The  95\% bootstrap interval for the double corrected variance estimate is the only estimator that covers the true variance (indicated by the vertical black lines) of model performance among the groups. }
    \Description{Each panel shows a different simulation scenario depending on whether the group sizes are equal or unequal and whether there are truly disparities or not in model performance among the groups. We show the histogram of 500 bootstrap samples and 95\% bootstrap intervals for the uncorrected (red), corrected (green), and double-corrected (blue) estimators of the between-group variance for one draw from the generative model.  The  95\% bootstrap interval for the double corrected variance estimate is the only estimator that covers the true variance (indicated by the vertical black lines) of model performance among the groups. }
        \label{fig:experiment3}
\end{figure*}

In summary, to obtain bootstrap intervals for the variance of $\mu$, we must account not only for the variance in the generative model, but also the additional variance that occurs due to the bootstrapping procedure itself. Figure \ref{fig:experiment3} shows bootstrap intervals and a histogram of bootstrap samples for the uncorrected, corrected, and double-corrected estimates of the between-group variance for one draw from the generative. Here we see that-- at least for this replicate-- by explicitly accounting for the additional variance induced by the bootstrap sampling process itself, we have once again successfully created an estimator that accurately captures between-group variance.

Extending beyond a single replicate, Table \ref{tab:coverage} shows the empirical coverage of 95\% bootstrap intervals across 1000 replicates. We see that applying the double-corrected variance estimator to create bootstrap intervals of the between-group variance has coverage similar to the nominal 95\% level.

\begin{table*}[ht]
\centering
\begin{tabular}{rrrr}
  \hline
 & uncorrected\_var & corrected\_var & double\_corrected\_var \\ 
  \hline
Equal Size; Equal Perf & 0.0 & 0.0 & 99.7 \\ 
  Unequal Size; Equal Perf & 0.0 & 0.0 & 99.3 \\ 
  Equal Size; Unequal Perf & 15.4 & 67.6 & 94.9 \\ 
  Unequal Size; Unequal Perf & 10.4 & 60.4 & 93.0 \\ 
   \hline
\end{tabular}
\caption{Empirical coverage of the 95\% bootstrap intervals over 1000 replicates. \label{tab:coverage}}
\Description{Empirical coverage of the 95\% bootstrap intervals over 1000 replicates.}
\end{table*}

\section{Real data example}\label{sec:realData}
We compare the variance and double-corrected variance estimators of group-wise model performance disparities in a model built from the Adult Income dataset\footnote{Downloaded from the UCI Machine Learning Repository at \url{https://archive.ics.uci.edu/ml/datasets/adult}.}, which is frequently used in algorithmic fairness research. It contains records of demographic and income data for 48,842 individuals from the 1994 census database. There are a total of 14 features, a weight column (\texttt{fnlwgt}), and a label column that indicates whether each person makes over \$50K a year. We split the data randomly into train and test sets by 70\%-30\%, and trained histogram-based gradient boosting classification trees\footnote{We used the scikit-learn implementation \url{https://scikit-learn.org/stable/modules/generated/sklearn.ensemble.HistGradientBoostingClassifier.html}.} using all 14 features of the dataset. The weight column is dropped as per prior research \cite{zhang2018mitigating, yurochkin2019training}. The resulting classifier has a 87.3\% accuracy on the test set.

\begin{figure*}[h]
    \centering
    \includegraphics{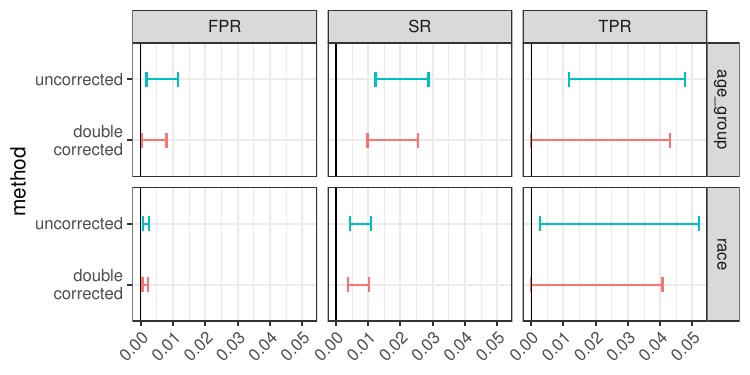}
    \caption{95\% bootstrap intervals for uncorrected and double-corrected estimates of variance over three performance metrics (FPR, SR, TPR) and two demographic features of the Adult income dataset. There are 8 age groups, 5 racial groups, and 40 intersectional groups. We see that we can come to different conclusions about whether there is between-group variability in model performance depending on which estimator we use. Notably, there are cases when the double corrected estimate says there could be no between-group variability when the uncorrected estimate says there is. }
    \Description{95\% bootstrap intervals for uncorrected and double-corrected estimates of variance over three performance metrics (FPR, SR, TPR) and two demographic features of the Adult income dataset. There are 8 age groups, 5 racial groups, and 40 intersectional groups. We see that we can come to different conclusions about whether there is between-group variability in model performance depending on which estimator we use. Notably, there are cases when the double corrected estimate says there could be no between-group variability when the uncorrected estimate says there is. }
    \label{fig:adult_income}
\end{figure*}

We are interested in how the model performs with respect to two demographic variables \textit{race} and \textit{age}. We divide age into 8 groups from 15 to 95, each spanning 10 years. The distribution of these two features are highly skewed. In the test set, across race, \textit{white} accounts for 86\% of the data, while \textit{American Indian and Eskimo} account for only 0.9\%. Across the 8 age groups, the smallest group, (85, 95], accounts for just 0.1\%, while the largest group, (25, 35], accounts for 27\%. We expect the presence of these small groups leads to increased statistical bias in estimated variance.

Figure~\ref{fig:adult_income} shows 95\% bootstrap intervals for the uncorrected and double corrected variance estimates. We calculated the intervals over three base model performance metrics, selection rate (SR), false positive rate (FPR), and true positive rate (TPR), since they are the base metrics that popular fairness criteria such as demographic parity and equalized odds depend on. The figure shows that the discrepancy between the uncorrected and double-corrected intervals is particularly large for TPR: the double-corrected intervals cover zero, while the uncorrected intervals do not. In other words, had we just used the uncorrected intervals, we would have concluded with high statistical certainty that there was a large disparity in model's TPR over both racial and age groups. However, after  accounting for uncertainty about TPR within each group, we conclude that the disparities are much lower, and we cannot statistically rule out zero disparities in false positives as a possibility.





\section{Conclusion}\label{sec:conclusion}
In this paper, we have demonstrated both theoretically and empirically that meta-metrics commonly used in the algorithmic fairness literature and in open-source tools designed for measuring group-wise model performance disparities are themselves statistically biased measurements. This occurs because existing meta-metrics fail to account for statistical uncertainty in the underlying base metrics, thus exaggerating between-group model performance disparities. In practice, this can cause misleading inferences about the extent of model performance disparities and erroneous conclusions about the relative degree of disparities between different sensitive variables. After identifying this problem, we propose a new estimator for evaluating between-group model performance disparities. This estimator builds upon the work of Cochran \cite{cochran1954combination} and Hedges and Olkin \cite{hedges2014statistical} by accounting for variability induced by bootstrap procedures. Our resulting double-corrected estimator offers a simple, conceptually clean, and easily implementable way to measure model performance disparities and associated uncertainty.

The method we have developed offers one approach to quantifying group-wise model performance disparities in the presence of many groups. To our knowledge, this is the first time this question has been studied in the context of algorithmic fairness. However, between-group variance estimation is well-studied in the statistics literature more generally. Common approaches include restricted maximum likelihood estimation of the variance parameter in random effects models \cite{harville1977maximum}. The literature on meta-analysis also contains many analogous methods. In that domain, the goal is to estimate variability in effect size across studies. There, $Y$ represents the effect size of a particular study, and $\sigma$ the standard error of that estimate. (See \cite{veroniki2016methods} for an excellent overview of methods available to estimate between-study effect size variance and its associated uncertainty in the context of meta-analysis and \cite{langan2019comparison} for a recent simulation study comparing methods.) Many of these approaches also offer solutions for quantifying model performance disparities across many groups. Future work should perform simulation studies comparing different methods for measuring between-group variability with parameters tailored to the specifics of typical problems encountered when evaluating model ``fairness."

Finally, model performance disparities as measured by meta-metrics are not dispositive of the presence or absence of algorithmic harms. While large disparities typically indicate that an issue needs further investigation, small measured disparities do not guarantee that the system is fair or free from adverse impacts. Just as it would be foolish to claim a computer system is completely secure or a data set is completely private, it is always possible that there are undiscovered vulnerabilities that our measurements have not uncovered. While we have presented an approach to more accurately measure model performance disparities, these measurements cannot tell us whether the observed disparities are acceptable or whether we have calculated disparities with respect to appropriate grouping variables. These judgments are subjective and require understanding of the context in which a model will be used.  Like all metrics, meta-metrics simply cannot capture the entirety of the impact of machine learning systems on people.

\bibliographystyle{ACM-Reference-Format}
\bibliography{bibliography}

\appendix
\section{Derivations}
Derivation of the single-corrected variance

\begin{eqnarray*}
\mathbb{E}[M_{\text{var}}(Y)] & = &  \mathbb{E} \left [\frac{1}{K-1}\sum_k ( Y_k - \bar{ Y})^2 \right ] \\
 & = &  \frac{1}{K-1} \mathbb{E}[\sum_k Y_k^2 - K \bar{Y}^2] \\ 
 & = & \frac{1}{K-1} \left(\sum_k \mu_k^2 + \sum_k \sigma^2_k  - K \mathbb{E}[\bar{Y}^2] \right)\\ 
 & = & \frac{1}{K-1} \left(\sum_k (\mu_k - \bar{\mu}  + \bar{\mu})^2 + \sum_k \sigma^2_k  - K \mathbb{E}[\bar{Y}^2] \right)\\ 
& = & \frac{1}{K-1} \left(\sum_k (\mu_k - \bar{\mu})^2  + K \bar{\mu}^2 + \sum_k \sigma^2_k  - K \mathbb{E}[\bar{Y}^2] \right)\\ 
& = &  m_{var}(\mu)  + \frac{1}{K-1} \left( K \bar{\mu}^2 + \sum_k \sigma^2_k  - K \mathbb{E}[\bar{Y}^2] \right)\\
& = & m_{var}(\mu)  + \frac{1}{K-1} \left( K \bar{\mu}^2 + \sum_k \sigma^2_k  - K (\bar{\mu}^2 + \frac{1}{K^2}\sum_k \sigma^2) \right)\\
& = & m_{var}(\mu)  + \frac{1}{K-1} \left( \sum_k \sigma^2_k  -  \frac{1}{K}\sum_k \sigma_k^2) \right)\\
& = & m_{var}(\mu) +  \frac{1}{K-1} \left( \frac{K-1}{K}\sum_k \sigma_k^2) \right)\\
& = & M_{\text{var}}(\mu) +  \left( \frac{1}{K}\sum_k \sigma_k^2 \right),
\end{eqnarray*}
\end{document}